\documentclass[twocolumn,showpacs,prb]{revtex4-1}
\usepackage[english]{babel} 
\usepackage{bm}
\usepackage{graphicx}
\usepackage{epsfig}
\usepackage{amsmath} 
\usepackage{color}
\usepackage{ulem}
\usepackage{amssymb}
\usepackage[colorlinks = true]{hyperref}

\newcommand{\eps}{\varepsilon} 
\renewcommand{\i}{\mathrm{i}}
\newcommand{\B}{\mathcal B}
\newcommand{\D}{\mathcal D}
\newcommand{\A}{\mathcal A}
\newcommand{\e}{\mathrm{e}}

\newcommand{\changes}[1]{#1}

\begin{document}

\title{High-frequency nonlinear transport and photogalvanic effects in two-dimensional topological insulators} 

\author{M.\,V.\,Durnev}
\author{S.\,A.\,Tarasenko}

\affiliation{Ioffe Institute, 194021 St.\,Petersburg, Russia}

\begin{abstract}
Excitation of a topological insulator by a high-frequency electric field of a laser radiation leads to a dc electric current in the helical edge channel whose direction and magnitude are sensitive to the radiation polarization and depend on the physical properties of the edge. We present an overview of theoretical and experimental studies of such edge photoelectric effects in two-dimensional topological insulators based on semiconductor quantum wells. First, we give a phenomenological description of edge photocurrents, which may originate from the photogalvanic effects or the photon drag effects, for edges of all possible symmetry.
Then, we discuss microscopic mechanisms of photocurrent generation for different types of optical transitions involving helical edge states. They include direct and indirect optical transitions within the edge channel and edge-to-bulk optical transitions.
\end{abstract}
\maketitle

\section{Introduction}

Topological insulators have emerged on the map of semiconductors as the class of narrow-gap materials with inverted band structure~\cite{Hasan2010, Zhang2011,Moore2010,Volkov1986}.
The band inversion originates from strong spin-orbit coupling in these materials and leads to the formation of helical (with spin-momentum locking)  
Weyl states on their surfaces~\cite{Volkov1985,Pankratov1987,Bernevig2006a,Kane2005a,Fu2007a}. The surface states in a topological insulator fully fill the band gap of the bulk material and are resistant to  
non-magnetic perturbations. Growing attention to topological insulators is fueled by a fundamental interest in solid-state spin physics
and the prospect of designing novel functional materials and electronic devices based on them~\cite{Ando2013}. 

Topological surface states have been theoretically predicted and subsequently observed in a number of three-dimensional (3D) binary and ternary V-VI 
compounds, such as Bi$_2$Se$_3$, Bi$_2$Te$_3$, Bi$_2$Te$_{3-x}$Se$_x$, Sb$_2$Te$_3$, and their solid solutions~\cite{Zhang2009, Xia2009, Ando2013}, II-VI compounds like
HgTe~\cite{Fu2007b,Brune2011,Kozlov2014}, etc. Examples of two-dimensional (2D) topological insulators with one-dimensional helical channels at the sample edges include
HgTe/CdHgTe~\cite{Bernevig2006,Konig2007} and InAs/GaSb~\cite{Liu2008,Krishtopenko2018} quantum wells (QWs) of certain thicknesses, and 1T' polytypes of transition metal dichalcogenide 2D crystals like WTe2~\cite{Qian2014,Fei2017}.  

Experimentally, surface states in 3D topological insulators are revealed by the angle and spin resolved photoemission spectroscopy~\cite{Xia2009,Hsieh2009}, in magneto-transport~\cite{Brune2011,Kozlov2014,Checkelsky2009,Analytis2010,Ren2011,Xia2013,Pan2014} 
and magneto-optical~\cite{Shuvaev2013b,Dziom2017} measurements. The properties of conducting edge channels in 2D topological insulators are primarily studied in local and non-local
transport measurements~\cite{Konig2007,Roth2009,Gusev2011,Hart2014,Ma2015,Tikhonov2015,Kononov2015,Kadykov2018}.  

\changes{Besides the measurements of the linear response, when the electric current $\bm j$ oscillates at the frequency $\omega$ of the applied electric field and scales linearly with the field amplitude $\bm E$, it is highly informative to study non-linear transport phenomena. In particular, excitation of a macroscopically homogeneous 
structure by the ac electric field may lead to a direct electric current $\bm j_{\rm dc}$ even in the absence of a dc bias. }
This is possible if the structure lacks the center of space inversion, e.g., due to the presence of a surface or an edge~\cite{ Magarill1979,Alperovich1982,Schmidt2015,Karch2011}. 
In the classical range of frequencies $\omega$, when the quantum of energy $\hbar \omega$ is much smaller than the mean 
kinetic energy of carriers, such effects are treated as nonlinear electron transport and may be included in the class of quantum ratchet phenomena~\cite{Reimann2002,Falko1989,Tarasenko2008,Drexler2013}. 
At higher frequencies, when mechanisms of the dc current formation originate from the asymmetry of optical transitions
in the $\bm k$ space, these effects are commonly named photogalvanic
effects~\cite{SturmanFridkin,IvchenkoGanichev}. The photogalvanic effects, discovered in 70s in bulk pyroelectric and gyrotropic crystals~\cite{Glass1974,Asnin78}, are extensively studied now in low-dimensional semiconductor 
structures providing an access to symmetry of the structures, optical selection rules, energy spectrum,
energy, momentum, and spin relaxation times of carriers, etc.~\cite{Ganichev2003,Bieler2005,Yang2006,Olbrich2009,Priyadarshi2012,Olbrich2013,Ganichev2014,Li2017,Mikheev2018}.
They are also used to optically inject spin~\cite{Bhat2005,Tarasenko2005,Zhao2005,Ganichev2006,Ganichev2009} and valley~\cite{Tarasenko2005,Karch2011b,Golub2011,Hartmann2011,Yuan2014,Linnik2014,LyandaGeller2015} currents controlled by light polarization.  
 
Photogalvanic spectroscopy of topological insulators~\cite{Hsieh2011,McIver2012,Olbrich2014,Dantscher2015,Shikin2016,Hamh2016,Galeeva2016,Pan2017,Huang2017,Kuroda2017,Dantscher2017,Plank2018} is of particular interest since the photocurrents associated with edges can be experimentally separated from the photocurrents stemming from the bulk of the 
structure. The edge and bulk contributions to the photoresponse have, as a 
rule, different polarization dependence. Moreover, for some experimental geometries, the bulk contribution vanishes due to symmetry arguments while the
edge contribution does not. Additional information on the nature of the photocurrent is obtained by studying the spectrum of the photocurrent excitation and 
the dependence of the photocurrent of the Fermi level. In particular, the photocurrent associated with helical edge states is also excited by radiation
with the photon energy $\hbar \omega$ smaller than the bulk band gap.
 
Here, we present an overview of theoretical and experimental study of edge photogalvanic effects in 2D topological insulators with the focus on HgTe/CdHgTe 
QW structures. We provide a phenomenological description of the circular and linear photogalvanic effects and the photon drag effect  
and analyze them for 2D topological insulators with all possible symmetry of the edges. Then we discuss microscopic mechanisms of the edge photocurrent 
generation for different types of optical transitions including optical transitions between the ``spin-up'' and ``spin-down'' branches, edge-to-bulk optical 
transitions, and indirect (Drude-like) optical transitions within the edge channels. Theoretical results are compared with available experimental data.

\section{Symmetry analysis} 

\begin{figure}[t]
\begin{center}
 \includegraphics[width=0.6\columnwidth]{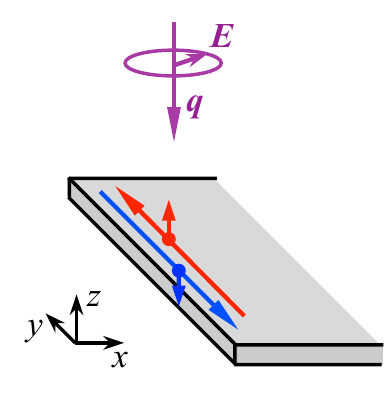}%
 \end{center}
 \caption{\label{fig1}
  Edge of a 2D topological insulator with a helical channel where the ``spin-up'' and ``spin-down'' electrons propagate in the opposite directions. The structure is excited by a polarized electromagnetic wave which leads to an edge photocurrent.}
\end{figure}

Consider a structure made of a 2D topological insulator with the edge along the $y$ axis, see Fig.~\ref{fig1}. The structure supports a pair of 
counterpropagating edge modes with the opposite spin projections. It is excited by an electromagnetic wave with the electric field given by
\begin{equation}
\bm E (\bm r ,t) = \bm E \exp(\i \bm q \cdot \bm r - \i \omega t) + \bm E^* \exp(-\i \bm q \cdot \bm r + \i \omega t)\:,
\end{equation}
where $\bm E$ is the electric field amplitude and $\bm q$ is the wave vector in the structure. The electromagnetic wave induces a dc electric current in the 
edge channel. 

The dc current emerges in the second order in the electric field amplitude, i.e., linear in the radiation intensity, and is described by the phenomenological 
equation~\cite{SturmanFridkin,IvchenkoGanichev}
\begin{equation}
\label{PGE}
j_{{\rm edge}}^{({\rm PGE})} = \sum_{\beta \gamma} L_{y\beta\gamma} (E_{\beta}E_{\gamma}^* + E_{\gamma}E_{\beta}^*) 
+ \sum_{\beta} C_{y\beta} \mathrm{i} [{\bf E} \times {\bf E}^*]_{\beta}\:,
\end{equation}
where $L$ and $C$ are the third-rank and second-rank tensors, and the indices $\beta$ and $\gamma$ run over the Cartesian coordinates $x$, $y$, and 
$z$.  The $L$ tensor describes the linear photogalvanic effect (LPGE), i.e., the photocurrent excited by linearly polarized ac electric field. The $C$ tensor 
stands for the circular photogalvanic effect (CPGE). This photocurrent is excited by elliptically or circularly polarized radiation and reverses its direction upon 
switching the sign of the photon helicity, since $\mathrm{i} [{\bf E} \times {\bf E}^*] = ({\bf q} / q) |\bm E|^2 P_{\rm circ}$, where $P_{\rm circ}$ is the degree of circular 
polarization. Both LPGE and CPGE are possible in systems lacking the center of space inversion: LPGE is allowed in piezoelectric structures~\cite{SturmanFridkin}, while CPGE
is allowed in gyrotropic structures~\cite{IvchenkoGanichev}, which follows from the symmetry of Eq.~\eqref{PGE}. Such a symmetry breaking naturally occurs at the edge of a 2D 
structure.  

\begin{figure*}[t]
 \includegraphics[width=0.99\textwidth]{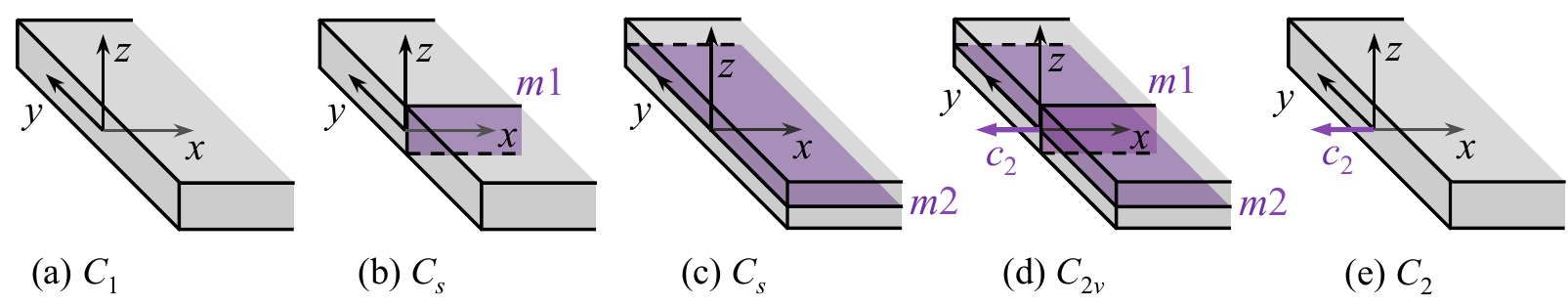}%
 \caption{\label{fig2}
Possible point groups and symmetry elements of semi-infinite quasi-two-dimensional structures. $m1$ and $m2$ are mirror planes and $c_2$ is the rotation
axis by 180 degree.}  
\end{figure*}

Elements of spatial symmetry present in the system may impose restrictions on the form of the $L$ and $C$ tensors.  
Generally, there are 5 types of semi-infinite quasi-two-dimensional crystalline structures. They are all illustrated in Fig.~\ref{fig2}. 

(i) 
A structure with an edge of the lowest possible symmetry is described by the $C_1$ point group which has no non-trivial symmetry elements, Fig.~\ref{fig2}a.
In such systems, all the components of the edge linear and circular photogalvanic tensors can be non-zero.

(ii)
A semi-infinite structure with an edge can have the mirror plane perpendicular to the edge, $m1 \parallel (xz)$ in Fig.~\ref{fig2}b. Such a system 
is described by the $C_s$ point group. This case is realized, e.g., in 2D topological insulators based on (001)-grown zinc-blende-type QWs with the edge along one of the crystallographic directions $\langle 110 \rangle$.

(iii)
A structure can have the mirror plane lying in the center of the quasi-two-dimensional layer, $m2 \parallel (xy)$ in Fig.~\ref{fig2}c. 
Such systems also belong to the $C_s$ point group but the mirror plane of the different orientation imposes different restrictions on the photogalvanic tensor
components. Examples are (110)-grown QWs with symmetric confinement potential and any edge orientation except $y \parallel [1\bar{1}0]$. 

(iv)
A structure can have both the $m1$ and $m2$ mirror planes, see Fig~\ref{fig2}d. Then, it is described by the $C_{2v}$ point group containing besides the two 
mirror planes the two-fold rotation axis $c_2$. This is the highest possible symmetry of a semi-infinite 2D system. Such a symmetry is realized in symmetric (110)-grown QWs with the edge $y \parallel [1\bar{1}0]$.
In fact, this symmetry corresponds to commonly used isotropic models of 2D topological insulators, e.g., Bernevig-Hughes-Zhang (BHZ) model~\cite{Bernevig2006}.

(v) Finally, a structure can possess the two-fold rotation axis $c_2$ but has neither $m1$ nor $m2$ mirror planes, Fig.~\ref{fig2}e. Examples of such 
structures of the $C_2$ point group are symmetric (001)-grown QWs with the edge $y \parallel \langle 100 \rangle$.

Non-zero components of the edge LPGE and CPGE tensors, $L$ and $C$, respectively, for all the point groups listed above are summarized in 
Tab~\ref{tab1}. The components describing the edge photocurrents at the normal incidence of radiation are in bold. It follows from the table that, in this 
geometry, the radiation induces both linear and circular photogalvanic currents. 

\begin{table*}
\centering
 \caption{\label{tab1} Non-vanishing components of the edge LPGE and CPGE tensors in structures of different symmetry.}%
 \begin{tabular}{|p{2cm}|p{7cm}|p{4cm}|l|}
 \hline
 Point group & Examples in zinc-blende-type QWs & edge LPGE tensor $L$ & edge CPGE tensor $C$  \\    
 \hline
 $C_1$ & low-symmetry edges  & all components & all components \\
 $C_s$ ($m1 \perp y$) & $y \parallel \langle 110 \rangle$ in asymmetric (001)-grown QWs  & $\bf{yxy}$, $yxz$  & $yx$, $\bf{yz}$   \\
 $C_s$ ($m2 \perp z$) & edges in symmetric (110)-grown QWs & $\bf{yxx}$, $\bf{yyy}$, $yzz$, $\bf{yxy}$  & $\bf{yz}$ \\
 $C_{2v}$ & $y \parallel [1\bar{1}0]$ in symmetric (110)-grown QWs & $\bf{yxy}$ & $\bf{yz}$ \\
 $C_{2}$ & $y \parallel \langle 100 \rangle$ in symmetric (001)-grown QWs & $\bf{yxy}$, $yxz$ & $yy$, $\bf{yz}$ \\
 \hline
  \end{tabular}
\end{table*} 

In addition to the photogalvanic effects, which typically predominate in the photoresponse of non-centro\-symmetric structures, a photocurrent may also originate
from the joint action of the electric and magnetic fields of the radiation or from the modulation of the ac electric field in the real space. These effects belong to
    the class of photoelectric phenomena caused by light pressure 
    (photon drag)~\cite{Barlow1954,Perel1973,Shalygin2006,Karch2010,Obraztsov2014,Shalygin2016}. In structures of low spacial symmetry, the direction of the photon drag current is
not necessarily locked to the photon wave vector $\bm q$. The electric current may even flow in the direction perpendicular to $\bm q$ and have a 
pronounced polarization dependence~\cite{Shalygin2006,Karch2010,Shalygin2016}. The edge photon drag effect is phenomenologically described by
\begin{multline}
\label{PDE}
j_{{\rm edge}}^{({\rm PDE})} = \sum_{\beta \gamma \delta} D_{y\beta\gamma\delta}^{(L)} q_{\beta} (E_{\gamma}E_{\delta}^* + E_{\delta}E_{\gamma}^*) + \\
+ \sum_{\beta} D_{y\beta\gamma}^{(C)} q_{\beta} \mathrm{i} [{\bf E} \times {\bf E}^*]_{\gamma}\:,
\end{multline}
where $D^{(L)}$ and $D^{(C)}$ are the tensors of the linear and circular photon drag effects. The results of a symmetry analysis of the photon drag effect for 
structures of different point groups are presented in Tab.~\ref{tab2}. Interestingly, in structures lacking the horizontal mirror plane $m2$ the edge 
current caused by the linear photon drag effect can emerge even at the normal incidence of radiation. The corresponding components of the $D^{(L)}$ tensor 
are in bold. 

\begin{table*}
\caption{\label{tab2}Non-vanishing components of the edge linear and circular PDE tensors in structures of different symmetry}%
  \begin{tabular}{|p{2cm}|p{9cm}|p{5cm}|}
\hline
 Point group & Linear PDE tensor $D^{(L)}$ & Circular PDE tensor $D^{(C)}$ \\    
 \hline
 $C_1$ & all components & all components \\
 $C_s$ ($m1 \perp y$) & $yyxx$, $yyyy$, $yyzz$, $yyxz$, $yxxy$, $yxyz$, $\bf{yzxy}$, $yzyz$ &  $yxx$, $yxy$, $yzx$, $yzy$, $yyz$   \\
 $C_s$ ($m2 \perp z$) &  $yxxx$, $yxyy$, $yxzz$, $yxxy$, $yyxx$, $yyyy$, $yyzz$, $yyxy$, $yzxz$, $yzyz$ & $yxz$, $yyz$, $yzx$, $yzy$ \\
 $C_{2v}$ & $yyxx$, $yyyy$, $yyzz$, $yyxz$, $yxxy$, $yzyz$ & $yzx$, $yzy$, $yyz$  \\
 $C_{2}$ & $yxxy$, $yxxz$, $yyxx$, $yyyy$, $yyzz$, $yyyz$, $\bf{yzxx}$, $\bf{yzyy}$, $yzzz$, $yzyz$ & $yxy$, $yxz$, $yyx$, $yzx$ \\
 \hline
  \end{tabular}
\end{table*}

\section{Microscopic mechanisms of edge photocurrents}

In this Section, we consider the mechanisms of photocurrent generation for different types of optical transitions involving topological edge states and present 
the results of microscopic calculations. We focus
on 2D topological insulators based on zinc-blende-type QWs, such as commonly studied HgTe/CdHgTe QWs. 

First, we briefly outline the electron structure of helical edge states in HgTe/CdHgTe QWs. These states are formed mainly from the electron-like 
$|E1,\pm 1/2 \rangle$ and heavy-hole $|H1, \pm 3/2 \rangle$ 
subbands~\cite{Bernevig2006}.  In the basis $|E1,+ 1/2 \rangle$, $|H1, + 3/2 \rangle$, $|E1,- 1/2 \rangle$, and $|H1, - 3/2 \rangle$, the electron states 
in symmetric (001)-grown QW (D$_{2d}$ point group) are described by the effective 4$\times$4 Hamiltonian~\cite{Durnev2016}
\begin{equation}
\label{eq:H_bulk}
\mathcal H_0(k_x,k_y) =
\left( 
\begin{array}{cccc}
\delta_E & {\rm i}\A k_+ & 0 & {\rm i} \gamma \e^{-2\i\theta} \\
-{\rm i} \A k_- & \delta_H & {\rm i} \gamma \e^{-2\i\theta} & 0\\
0 & -{\rm i} \gamma \e^{2\i\theta} & \delta_E & - {\rm i} \A k_- \\
-{\rm i} \gamma \e^{2\i\theta} & 0 & {\rm i} \A k_+ & \delta_H
\end{array}
\right) \: ,
\end{equation}
where $\bm k = (k_x, k_y)$ is the in-plane electron wave vector, $k = |\bm k|$, $k_\pm = k_x \pm \i k_y$, 
$\delta_E = \delta_0 - (\B+\D)k^2$, $\delta_H = - \delta_0 + (\B-\D)k^2$; $\A$, $\B$, $\D$, $\gamma$, and {$\delta_0$} are the real-valued band-structure 
parameters. 

The parameter $\delta_0$ describes the band gap and defines whether the system is in the trivial ($\delta_0 >0$ at $\B < 0$) or non-trivial ($
\delta_0 < 0$ at $\B < 0$) topological phase~\cite{Bernevig2006}. The parameter $\gamma$ describes 
\changes{ the mixing of the $|E1, \pm 1/2 \rangle$ and $|H1, \mp 1/2 \rangle$
subbands }
at $\bm k = 0$ due to interface inversion asymmetry and bulk inversion asymmetry in the QW  structure~\cite{Tarasenko2015,Dai2008,Konig2008,Winkler2012}.
Its experimental value is not well determined yet. Theoretical estimations for the HgTe/CdHgTe QWs of the close-to-critical thickness yield  $\sim 2$~meV in the $\bm k$$\cdot$$\bm p$ model~\cite{Konig2008,Winkler2012} and $\sim 5$~meV in the atomistic (tight-binding and pseudopotential) approaches~\cite{Tarasenko2015,Dai2008}.

The Hamiltonian~\eqref{eq:H_bulk} is written in the in-plane coordinate frame $x \parallel ([100] \cos \theta - [010] \sin \theta)$, 
$y \parallel ([010] \cos \theta + [100] \sin \theta)$, and $z \parallel [001]$, where $[100]$,  $[010]$, and $[001]$ are the cubic axes, and $\theta$ is the
angle describing the edge orientation. Below, we 
use the following set of the band-structure parameters: $\A = 3.6$~eV$\cdot$\AA, $\B = -68$~eV$\cdot$\AA$^2$, $\D = -51$~eV$\cdot$
\AA$^2$~\cite{Konig2008}, $\gamma = 5$~meV~\cite{Tarasenko2015}, and $\delta_0 = -10$~meV, which corresponds to the topological gap 
of about $20$~meV. 

The energy spectrum and wave functions of electron states in the topological insulator with the edge at $x=0$
are found from the Schr\"{o}dinger equation
\begin{equation}
\mathcal H_0(- \i \nabla_x, k_y) \Psi = \varepsilon \Psi
\end{equation}
with a boundary condition for the wave function $\Psi$. We use the open boundary condition, i.e., $\Psi = 0$ 
at $x = 0$~\cite{Durnev2016,Sonin2010,Scharf2012,Klipstein2015}.  Boundary conditions of more 
general form were analyzed in Ref.~\cite{Enaldiev2015}.

\begin{figure}[t]
\begin{center}
 \includegraphics[width=0.85\columnwidth]{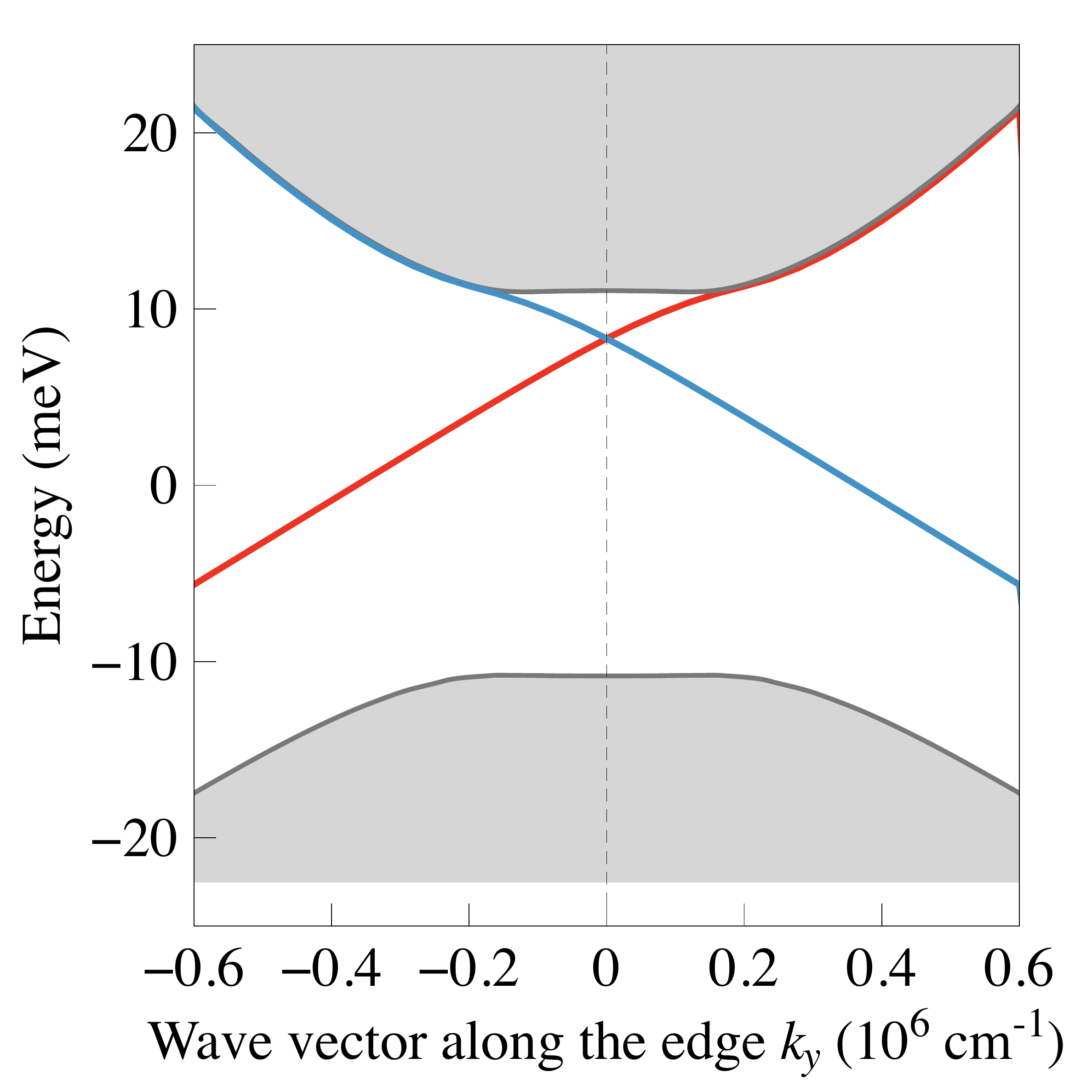}%
 \end{center}
 \caption{\label{fig3}
Energy spectrum of a semi-infinite structure made of HgTe/CdHgTe-based 2D topological insulator with the band gap of $\approx 20$~meV.
The branches of helical edge states are blue and red colored.} 
\end{figure}

Figure~\ref{fig3} shows the energy spectrum of a topological-insulator structure calculated for the above parameters. 
The spectrum contains a pair of spin-polarized branches (shown by blue and red curves) which are
in the band gap of the bulk QW states and correspond to the helical edge modes. The whole energy spectrum is asymmetric with respect 
to positive and negative energy due to electron-hole asymmetry introduced by the parameter $\D$ in the Hamiltonian~\eqref{eq:H_bulk}.
In particular, the edge-state branches are shifted from the band gap center. At small wave vector along the edge $k_y$, the edge states have linear 
dispersion and can be described by the effective $2 \times 2$ Weyl Hamiltonian
\begin{equation}
H_{{\rm edge}} = \hbar v_0 \sigma_z k_y \,,
\end{equation}
where $v_0$ is the edge-state velocity and $\sigma_z$ is the Pauli matrix in the pseudospin-$1/2$ space
\changes{of the edge-state wave functions at $k_y = 0$, $\psi_{0, \pm 1/2}$. } 
At large $k_y$, the dispersion
of the edge states deviate from the linear law. The Hamiltonian~\eqref{eq:H_bulk}
and the corresponding spectrum in Fig.~\ref{fig3} are obtained in the 4-subband model. 
This model does not take into account the complex structure of the valence band in the commonly studied 8-nm-wide HgTe/CdHgTe QWs. In fact, the 
valence band is strongly affected by the closely lying $H2$ subband not included in the model, which leads to a nonmonotonic dispersion of the hole states with side maxima~\cite{Raichev2012,Krishtopenko2018,Minkov2016}.

Now, we discuss the interaction of electromagnetic waves with the topological insulator and the origin of edge photocurrent generation.
Depending on the photon energy and the position of the Fermi level in the structure, edge photocurrents are contributed by different types of optical 
transitions. 

\subsection{Direct optical transitions between spin branches}\label{sec:inter-branch}

\begin{figure}[htpb]
\begin{center}
\includegraphics[width=0.2\textwidth]{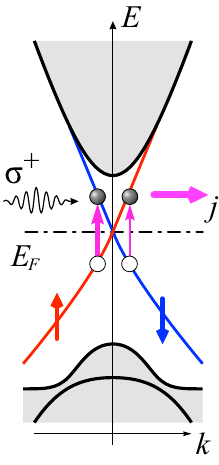}
\end{center}
\caption{\label{fig:direct} Edge photocurrent caused by direct optical transitions between the ``spin-up'' and ``spin-down'' edge states in 2D topological insulator. Spin-dependent asymmetry of the optical transitions induced by polarized radiation leads to an imbalance of electrons between the spin states giving rise to a spin polarization and a direct electric current. }
\end{figure}

Direct optical transitions between the edge states with the pseudospin projections $s = \pm 1/2$, see Fig.~\ref{fig:direct}, take place when 
$\hbar \omega > 2 | \eps_F|$, where $\eps_F$ is the Fermi energy counted from the Weyl point~\cite{Dora2012,Artemenko2013,Durnev2018}. If, additionally, the inequality   
$\hbar \omega < \eps_c - \eps_F, \eps_F - \eps_v$ is satisfied, where $\eps_c$ and $\eps_v$ are the energies of the conduction-band bottom
and the valence-band top in the QW bulk, respectively, these are the only possible optical transitions in a pure structure. 

Because of the selection rules to be discussed below, the optical transitions $|-k_y, +1/2\rangle \rightarrow |-k_y, -1/2\rangle$ and 
$|k_y, -1/2\rangle \rightarrow |k_y, +1/2\rangle$ induced by polarized radiation occur at different rates. This is illustrated in Fig.~\ref{fig:direct} by vertical 
arrows of different thicknesses. The asymmetry of the optical transitions results in an imbalance of electrons between the spin-momentum locked $+1/2$
and $-1/2$ branches giving rise to an electron spin polarization and a direct electric current. After an optical pulse, the spin polarization and the edge
current decay with the characteristic time determined by spin-flip backscattering processes which are very rare. Under cw excitation, the decay is 
compensated by the optical generation of new spin-polarized electrons, and the photocurrent reaches its steady-state magnitude.

The optical transitions between the spin branches occur due to the interaction of edge carriers with the electric field $\bm E(t)$~\cite{Durnev2018} or magnetic field $\bm B(t)$~\cite{Dora2012,Artemenko2013} of the incident electromagnetic wave. The Hamiltonian of the electro-dipole electron-photon interaction has the form 
\begin{equation}\label{eq:Ham_ed}
\mathcal H_{\rm edge}^{(\bm E)} = -\bm d \cdot \bm E(t) \,,
\end{equation}
where $\bm d$ is the electric dipole operator.   
\changes{
$\mathcal H_{\rm edge}^{(\bm E)}$ can be also rewritten in the equivalent form as $-(e/c) \bm v \cdot \bm A(t)$, where
$e$ is the electron charge, $c$ is the speed of light, $\bm v$ is the velocity operator, and $\bm A(t)$ is the vector potential 
of the electromagnetic wave whose dependence on the coordinate is neglected.
}
Microscopic theory based on the Hamiltonian~\eqref{eq:H_bulk} 
shows that the inter-branch matrix elements of $\bm d$ is non-zero~\cite{Durnev2018}. In the basis of the edge-state wave functions 
at $k_y = 0$, $\psi_{0, \pm 1/2}$, the components of the electric dipole operator at small $k_y$ read
\begin{eqnarray}\label{dxdy}
d_x &=&  (\sigma_y \cos 2\theta - \sigma_x \sin 2\theta) D_{1} k_y \:, \nonumber \\
d_y &=&  (\sigma_x \cos 2\theta + \sigma_y \sin 2\theta) D_{2} k_y \:, 
\end{eqnarray}
where $\sigma_x$ and $\sigma_y$ are the Pauli matrices, and $D_1$ and $D_2$ are real parameters. Both $d_x$ and $d_y$ vanish at $k_y = 0$
since the time reversal symmetry does not allow a lifting of the Kramers degeneracy of the states $|0, +1/2\rangle$ and $|0, -1/2\rangle$ by an electric  
field. 

We also note that the electro-dipole optical transitions are possible due to the lack of space inversion center in the QW structure and the parameters $D_1$ and $D_2$ scale with constant of the subband mixing $\gamma$ in the Hamiltonian~\eqref{eq:H_bulk}. 
\changes{Indeed, at zero $\gamma$ the Hamiltonian~\eqref{eq:H_bulk} assumes the block-diagonal form and the edges states
with $s =\pm 1/2$ stemming from these blocks are decoupled.  Numeric calculations show that $|D_1/e| \approx 7\times10^{-13}$~cm$^2$ and $|D_2 /e| \approx 1.5\times10^{-12}$~cm$^{2}$ for the band-structure parameters listed above~\cite{Durnev2018}. Deriving analytical expressions for $D_{1,2}$ in terms of the band-structure parameters is a challenging task.}
In isotropic approximations,  
which corresponds to the effective $C_{2v}$ point group for the QW structure with an edge, the components  $d_x$ and $d_y$ vanish at any $k_y$.

The Hamiltonian of the magneto-dipole interaction (the Zeeman Hamiltonian) in the basis of the wave functions $\psi_{0, \pm 1/2}$ has the form
\begin{equation}
\label{eq:Ham_md}
\mathcal H_{\rm edge}^{(\bm B)} = - \bm{\mu} \cdot \bm{B}(t)  = \frac{\mu_B}{2} \sum \limits_{\alpha, \beta = x,y,z} g_{\alpha \beta} \sigma_\alpha B_\beta(t) \:,
\end{equation}
where $\bm{\mu}$ is the magnetic dipole operator, $\mu_B$ is the Bohr magneton, and $g_{\alpha \beta}$ are the $g$-factor tensor components. 
The in-plane components are given by~\cite{Durnev2016}
\begin{eqnarray}\label{g_all}
g_{xx} &=& g_{1} \cos^2 2\theta + g_{2} \sin^2 2\theta\:, \nonumber \\
g_{yy} &=& g_{1} \sin^2 2\theta + g_{2} \cos^2 2\theta\:, \nonumber \\
g_{xy} &=& g_{yx} = \frac12 \left( g_{1} - g_{2} \right) \sin 4 \theta\:,
\end{eqnarray}
where $g_{1}$ and $g_{2}$ are two independent $g$-factors. The Zeeman gap  in the edge-state spectrum opened by the out-of-plane magnetic field is theoretically studied in Ref.~\cite{Durnev2016}   

With the account for both the electro-dipole and the magneto-dipole mechanisms of electron-photon interaction,
the optical transitions $|k_y, -s\rangle \rightarrow |k_y, s\rangle$ induced by electromagnetic wave polarized in the QW plane 
are described by the matrix elements
\begin{equation}
\label{Mopt}
M_{s\,, -s} (k_y)= - \bm{d}_{s\,, -s} \cdot \bm{E} - \bm{\mu}_{s\,, -s} \cdot \bm{B} \,,
\end{equation}
where $\bm{d}_{s\,, -s}$ and $\bm{\mu}_{s\,, -s}$ are the matrix elements of the electric dipole and magnetic dipole operators,
respectively, $\bm E$ and $\bm B$ are the amplitudes of the electric and magnetic fields of the radiation related by
$\bm{B} = n_{\omega} \, \bm{o} \times \bm{E}$, $n_{\omega}$ is the refractive index of the medium, and $\bm{o}$ is the unit vector 
along the radiation propagation direction $\pm z$. 
\changes{It follows from Eqs.~\eqref{dxdy} and~\eqref{g_all} that, at small $k_y$, the matrix elements are given by 
\begin{eqnarray}
M_{\pm 1/2 , \mp 1/2}(k_y) = \pm \i \e^{\mp 2\i \theta} (D_1 E_{x} \pm \i D_2 E_{y}) k_y \hspace{1.2cm} \\
+ \frac{\mu_B}{2} \left[ \frac{g_1 + g_2}{2} (B_{x} \mp \i B_{y}) + \frac{g_1 - g_2}{2} \e^{\mp 4 \i \theta}(B_{x} \pm \i B_{y}) \right] \:. \nonumber
\end{eqnarray}
} 

The photocurrent originating from asymmetric corrections to the electron distribution function in the relaxation time approximation
is given by~\cite{Ganichev2003,Durnev2018} 
\begin{eqnarray}\label{eq:jy} 
j_y &=& \frac{2 \pi e }{\hbar}  \sum_{k_y \, s} \tau_p [ v_{k_y\, -s} - v_{k_y s} ] \left| M_{-s, s} \right|^2  \nonumber \\
&\times& [f(\eps_{k_y s})-f(\eps_{k_y \, -s})] \delta(\eps_{k_y \, -s} - \eps_{k_y s} - \hbar \omega) \,,
\end{eqnarray} 
where $e$ is the electron charge, $\tau_p$ is the relaxation time, $v_{k_y s} = (1/\hbar) d \eps_{k_y s} / d k_y$ and $\eps_{k_y s}$  are the electron velocity and energy, respectively, 
and $f(\eps)$ is the Fermi-Dirac distribution function. The time $\tau_p$ in Eq.~\eqref{eq:jy} is the spin-flip scattering time of thermalized 
electrons at the Fermi level if electron thermalization within the edge channel, which is governed by electron-electron collisions and electron-phonon
interaction, is faster than spin-flip scattering. In the opposite case, if electron thermalization is inefficient, $\tau_p$ is the spin-flip scattering time of hot 
carriers.

The optical orientation of edge electrons and the photocurrent sensitive to the photon helicity emerge in the electro-dipole approximation. 
The calculation of the photocurrent~\eqref{eq:jy} with the 
matrix elements of the optical transitions determined by the electric dipole operator~\eqref{dxdy} yields~\cite{Durnev2018} 
\begin{equation}\label{j_circ_final}
j_y^{({\rm CPGE})} = - \frac{4e \tau_p v_0 w}{\hbar \omega} \frac{D_1 D_2}{D_1^2 + D_2^2} I P_{\rm circ} o_z \:,
\end{equation}
where $I = c n_{\omega} | \bm E|^2/(2 \pi)$ is the radiation intensity, $c$ is the speed of light, $w$ is the absorption width of the edge channel for circularly polarized radiation 
calculated in the electro-dipole approximation, 
\begin{equation}
w = \frac{\pi \omega^3 (D_1^2 + D_2^2) \Delta f}{4 c n_{\omega} \hbar v_0^3} \:,
\end{equation} 
and $\Delta f = f(-\hbar\omega/2) -  f(\hbar\omega/2)$. Equation~\eqref{j_circ_final} describes the CPGE current which corresponds to the 
phenomenological parameter $C_{yz}$ in Eq.~\eqref{PGE} and Tab.~\ref{tab1}.

For linearly polarized radiation, spin-dependent asymmetry in the optical transitions and the corresponding electric current
emerge due to the interference of the electro-dipole and magneto-dipole transitions.
This photocurrent has the form~\cite{Durnev2018} 
\begin{eqnarray}\label{j_lin_final}
j_y^{({\rm PDE})} =  \left[A + B \left( \left| e_x \right|^2 - \left| e_y \right|^2 \right)\right] q \cos 2\theta \, I \nonumber \\ 
+ B ( e_x e_y^* + e_y e_x^* ) q \sin 2\theta \,  I \:,
\end{eqnarray}
where 
\begin{eqnarray}\label{eq:Q}
A &=& - \frac{4e c \tau_p v_0^2 \, w }{\hbar \omega^3} \frac{\mu_B (D_1 g_2 - D_2 g_1)}{D_1^2 + D_2^2} \:, \nonumber \\
B &=& -  \frac{4e c \tau_p v_0^2 \, w}{\hbar \omega^3} \frac{\mu_B (D_1 g_2 + D_2 g_1)}{D_1^2 + D_2^2}  \:,
\end{eqnarray}
and $\bm e  = \bm E /E$ is the unit vector of the radiation polarization.
The photocurrent~\eqref{j_lin_final} depends on the orientation of the edge with respect to the crystallographic axes and the radiation polarization plane with respect to the edge. It may also
appear when the sample is excited by unpolarized radiation. The photocurrent~\eqref{j_lin_final} belongs to the class of linear photon drag effect, see first term in Eq.~\eqref{PDE}. For $\theta = 0$, which corresponds to the edge along $[010]$ and the $C_2$ point group of the structure, Eq.~\eqref{j_lin_final} shows that the linear PDE is described by the components $D_{yzxx}^{(L)}$ and $D_{yzyy}^{(L)}$. For $\theta = \pi/4$, which corresponds to the edge along $[110]$ and the $C_s$ point group of the structure, the linear PDE is described by the component $D_{yzxy}^{(L)}$. Both cases are in agreement with the symmetry analysis of the photon drag effect summarized in Tab.~\ref{tab2}.

\begin{figure}[htpb]
\includegraphics[width=0.45\textwidth]{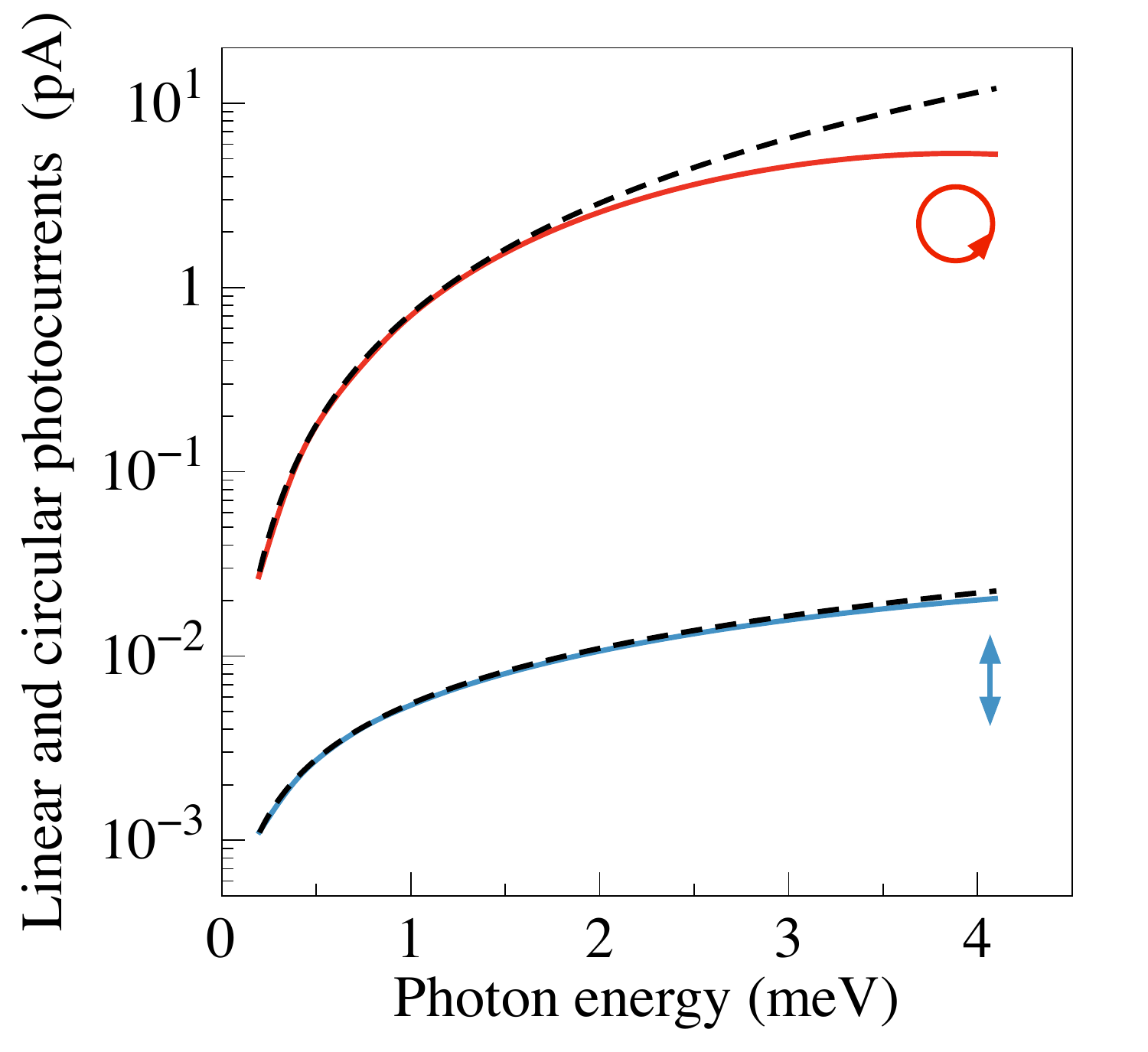}
\caption{\label{fig:current_interbranch} Edge photocurrents originating from the circular photogalvanic effect and the linear photon drag effect
as a function of the incident photon energy. The dependences are
calculated for the parameters of HgTe/CdHgTe-based 2D topological insulators, the momentum relaxation time
$\tau_p = 20$~ps, the refractive index $n_{\omega} = 3$, and the radiation intensity $I=1$~W/cm$^2$. The linear photocurrent is calculated for $\theta = 0$ and $\bm e \parallel y$. Solid curves show the results of numerical calculations, dashed curves are plotted after analytical Eqs.~\eqref{j_circ_final} and~\eqref{j_lin_final}.
}
\end{figure}

Figure~\ref{fig:current_interbranch} shows the magnitudes of the edge photocurrents originating from the circular photogalvanic effect and the linear photon drag effect in a HgTe/CdHgTe topological insulator as a function of the photon energy $\hbar\omega$. The dependences are calculated for zero temperature, the Fermi level lying at the Weyl point, and the relaxation time $\tau_p = 20$~ps estimated from the experiments on low-frequency edge conductivity~\cite{Dantscher2017}. Solid curves present the results based on numerical calculations of the matrix elements of the electron-photon interaction at finite $k_y$. Dashed curves show the low-energy analytical results plotted after Eqs.~\eqref{j_circ_final} and~\eqref{j_lin_final}. For the radiation intensity 1~W/cm$^2$, the photon energy $1$~meV, and the momentum relaxation time presented above, the photogalvanic and the photon drag currents at the normal incidence of radiation are in pA and fA ranges, respectively.   

\subsection{Edge-to-bulk optical transitions}\label{sec:edge-to-bulk}

\begin{figure}[htpb]
\begin{center}
\includegraphics[width=0.2\textwidth]{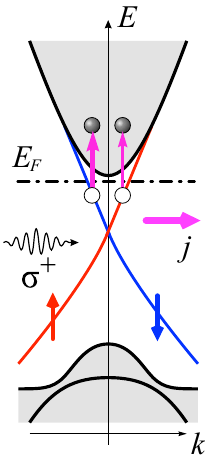} 
\end{center}
\caption{\label{fig:photoion} Photocurrent in the edge channel caused by edge-to-bulk optical transitions in 2D topological insulator. The transitions from 
the ``spin-up'' and ``spin-down'' branches induced by circularly polarized radiation have different rates, which results in different depopulations of the spin 
branches and produces an electric current.
} 
\end{figure}

Another class of optical transitions, which can occur at edges of 2D topological insulators, are edge-to-bulk transitions leading to photoionization of the 
edge channel~\cite{Dantscher2017,Artemenko2013,Kaladzhyan2015,Magarill2016}.
These are the transitions between one-dimensional edge states $| k_y, s\rangle$ and delocalized 2D states $|\bm k, s'\rangle$ belonging to the
conduction or the valence band, where $\bm k$ is the 2D wave vector.  The optical transitions from the edge channel to the conduction band take place 
if $\hbar \omega > \eps_c - \eps_F$ while the transitions from the valence band to the edge channel come into play at $\hbar\omega > \eps_F - \eps_v$.
The edge-to-band optical transitions induced by circularly polarized radiation and the corresponding mechanism of the generation of circular 
photocurrent in the edge channel are illustrated in Fig.~\ref{fig:photoion}.

Edge-to-bulk optical transitions are allowed in the electro-dipole approximation already in the isotropic BHZ model~\cite{Kaladzhyan2015}. Therefore, they predominate if the photon energy is large enough to throw up electrons from the 
edge states to the conduction band (or from the valence band to the edge states above the Fermi level). 

In the isotropic BHZ model, the optical transitions between the edge and conduction-band states occur with the pseudospin conservation.
However, the rates of such transitions from the $|k_y, +1/2 \rangle$ and $|- k_y, -1/2 \rangle$ states do not coincide with each other for circularly polarized 
radiation. This is similar to spin-dependent spin-conserving optical transitions between electron subbands in quantum wells~\cite{Ganichev2003}. The difference in the photoionization rates is shown in Fig.~\ref{fig:photoion} by vertical arrows of different thicknesses. 
Microscopically, the selection rules here are related to the fact that the edge states contain non-equal portions of the $E1$ and $H1$ states
due to electron-hole asymmetry described by the parameter $\D$ in the Hamiltonian~\eqref{eq:H_bulk}.

\changes{
The rates of the photoionization with the pseudospin conservation are given by the Fermi rule 
\begin{multline}
\label{generation}
g_{k_y s} = \frac{2 \pi}{\hbar} \sum_{k_x} |M_{s s}(k_x, k_y)|^2 
[f(\varepsilon_{k_y s}) - f(\varepsilon_{k_x k_y s}^c)] \times \\
\times \delta(\varepsilon_{k_x k_y s}^c - \varepsilon_{k_y s} -\hbar\omega) \,, 
\end{multline}
where $M_{s s}(k_x, k_y)$ is the matrix element of the optical transitions from the edge states $| k_y ,s \rangle$ to the conduction-band states $|k_x k_y ,s \rangle$, $\varepsilon_{k_x k_y s}^c$ is the energy of the conduction-band electrons, and $k_x$ is the wave vector of the conduction-band electrons in the direction normal to the edge.  
}

 The relative difference of the photoionization rates from the $|k_y, +1/2 \rangle$ and $|- k_y, -1/2 \rangle$ states, $g_{k_y+1/2}$ and $g_{-k_y -1/2}$, respectively, is proportional to the helicity of incident photons $P_{\rm circ}$ and may be presented in the form
\begin{equation}
\label{K}
\frac{g_{k_y +1/2} - g_{-k_y -1/2}}{g_{k_y +1/2} + g_{-k_y -1/2}} = K P_{\rm circ} o_z \,.
\end{equation}
The dimensionless coefficient $K$ at $k_y=0$ is given by $2 \B \D/(\B^2 + \D^2)\approx 0.96$, Ref.~\cite{Kaladzhyan2015}.
Numerical calculations show that the coefficient $K$ has only weak dependence on $k_y$ and the photon energy $\hbar\omega$~\cite{Dantscher2017}.

To calculate the photocurrent originating from spin-selective photoionization of the edge channel we consider that the momentum relaxation time of bulk
carriers is much shorter than that of edge carriers and neglect the photocurrent in the bulk states. We also assume that spin relaxation of bulk carriers is 
fast and photoionized electrons get unpolarized before they are trapped back on the helical edge states. In that case, the edge photocurrent is given by~\cite{Dantscher2017}
\begin{equation}
\label{jx}
j_y^{(\rm CPGE)}  = - \frac{e \tau_p v_0 w}{\hbar\omega} K I P_{\rm circ} o_z \,,
\end{equation}
where $w = (\hbar \omega/I) \sum_{k_y} (g_{k_y +1/2} + g_{-k_y -1/2})$ is the absorption width of the edge channel for circularly polarized radiation.

\begin{figure}[htpb]
\begin{center}
\includegraphics[width=0.45\textwidth]{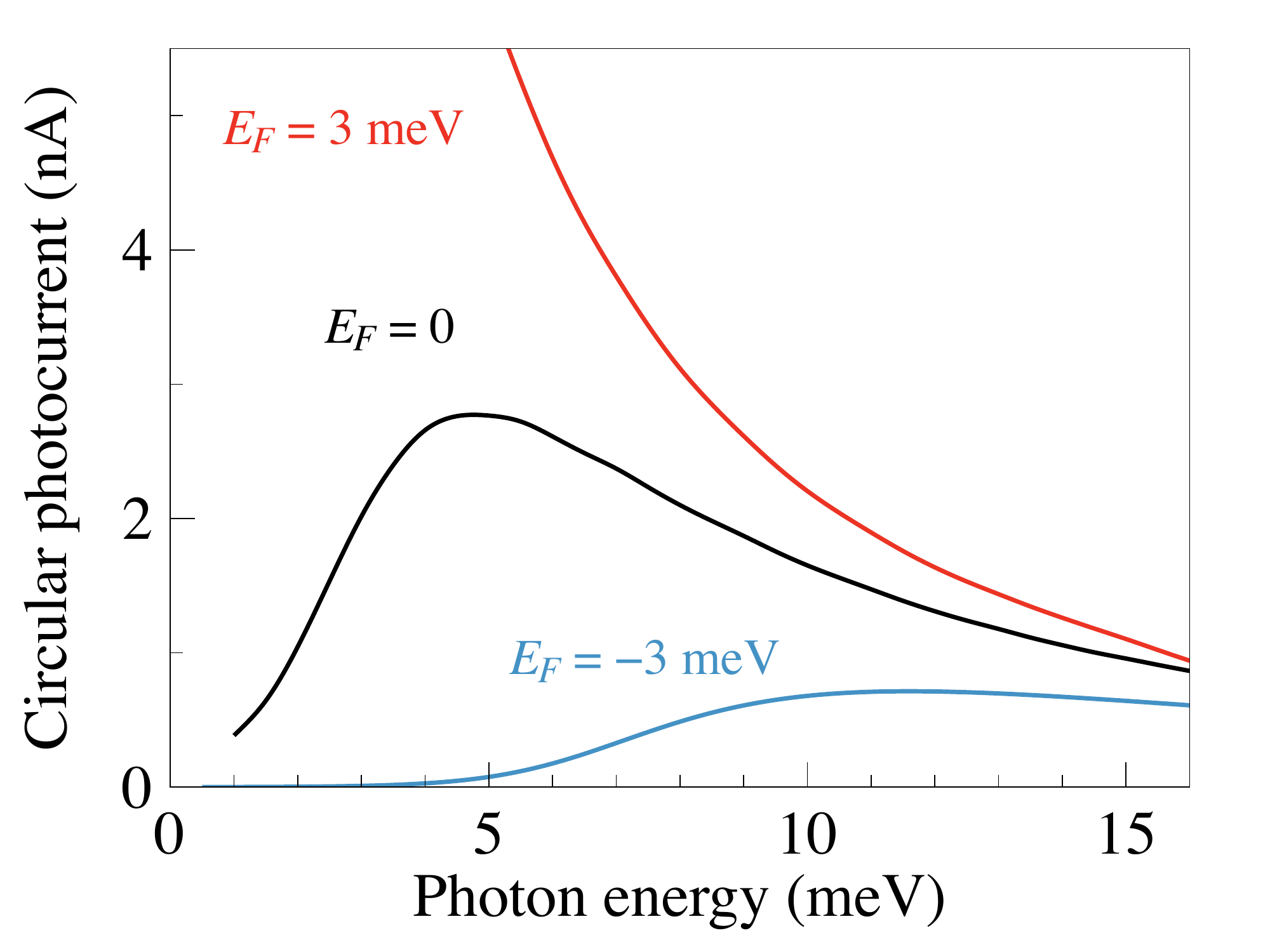}
\end{center}
\caption{\label{fig:j_photoion} Spectral dependence of the circular edge photocurrent that emerges due to optical transitions from edge states
to conduction-band states. Different curves correspond to different positions of the Fermi level counted from the Weyl point. The curves 
are calculated for the parameters of HgTe/CdHgTe-based 2D topological insulators, the momentum relaxation time $\tau_p = 20$~ps, 
the refractive index $n_{\omega} = 3$, and the radiation intensity $I=1$~W/cm$^2$.
}
\end{figure}

The excitation spectrum of the edge photocurrent for three different positions of the Fermi level is shown in Fig.~\ref{fig:j_photoion}.
For the Fermi level lying inside the bulk gap, the photocurrent has a threshold dependence on $\hbar \omega$. With the increase of $\omega$ 
the photocurrent rises, reaches its maximum, and then decreases. The decrease of the photocurrent at large $\omega$ is caused by
the reduction of the photon density at a fixed radiation intensity $I$ and the reduction of the probability of the edge-to-band optical transitions. 
The latter stems from the weak overlap between the wave functions of edge and bulk states involved at large $\omega$. 

\subsection{Indirect optical transitions in the edge channel}

\begin{figure}[htpb]
\begin{center}
\includegraphics[width=0.4\textwidth]{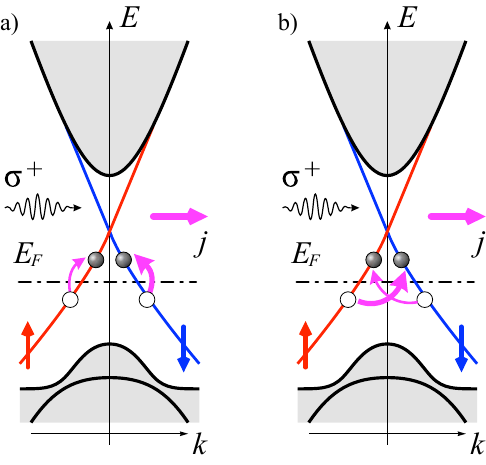}
\end{center}
\caption{\label{fig:drude} Edge photocurrent due to indirect (assisted by electron scattering from static defects or phonons) optical transitions in the edge channel. (a) Indirect optical transitions within the spin branches, (b) indirect spin-flip optical 
transitions.
}
\end{figure}

At small photon energy, when neither of the direct optical transitions considered above are possible, radiation can still be absorbed in an edge channel and leads to a photocurrent~\cite{Entin2016}. In this case, the absorption of photons occurs as the result of indirect (Drude-like) optical transitions assisted by electron scattering from static defects or phonons to satisfy the energy and momentum conservation, Fig.~\ref{fig:drude}. The photogalvanic effect for this type of transitions in quantum wells was considered in Ref.~\cite{Tarasenko2007}.

Indirect optical transitions are theoretically treated as virtual processes via intermediate states, which can belong to the edge channel or the 2D conduction or valence subbands. The compound matrix element of the transitions 
with the initial $|k_y, s \rangle$ and final $|k_y', s' \rangle$ states assisted by elastic scattering has the form 
\begin{equation}
\label{M}
M_{k_y' s', k_y s} = \sum \limits_{m} \left( \frac{V_{k_y' s',m} R_{m, k_y s}}{\eps_{k_y s} - \eps_{m} + \hbar \omega} + \frac{R_{k_y' s',m} V_{m, k_y s}}{\eps_{k_y s} - \eps_{m}} \right)\:, 
\end{equation}
where the index $m$ runs over intermediate states, $R_{m, k_y s}$ is the matrix element of electron-photon interaction, and $V_{k_y' s', m}$ is the matrix element of electron scattering. 

There are two types of indirect optical transitions in edge channels: (a) transitions within the same spin branches, i.e., 
with $s' = s$ in Eq.~\eqref{M}, and (b) inter-branch (spin-flip) transitions, i.e., with $s' = - s$. They are sketched in Fig.~\ref{fig:drude}a and Fig.~\ref{fig:drude}b, respectively. 

(a) For optical transitions with the pseudospin conservation, Fig.~\ref{fig:drude}a, intermediate states lie either in the same edge branch or in the 2D conduction or valence subbands. The matrix element of the virtual transitions with intermediate states in the same branch is given by
\begin{equation}\label{M_via_edge}
M_{k_y' s, k_y s}^{({\rm e})} = \i \frac{e (v_{k_y s} - v_{k_y' s}) E_y}{\hbar \omega^2} V_{k_y' s, k_y s}  \:.
\end{equation}
Since the dispersion of the edge states is almost linear and the velocity $v_{k_y s}$ has only weak dependence on $k_y$, this contribution is small. 
The matrix element of the virtual transitions via conduction-band or valence-band 2D states
can be estimated as 
\begin{equation}\label{M_via_edge}
M_{k_y' s, k_y s}^{({\rm b})} \sim \frac{e \A |\bm E|}{\hbar \omega \delta_0} V_{E1} \:,
\end{equation}
where $V_{E1}$ is the matrix element of scattering.
It is assumed here that the photon energy $\hbar\omega$ is much smaller than the energy separation between the Fermi level and 2D states which is of the order of the bulk band gap. Similarly to the direct edge-to-bulk transitions considered in Sec.~\ref{sec:edge-to-bulk}, the matrix elements  $M_{k_y ' +1/2,k_y +1/2}^{({\rm b})}$ and $M_{-k_y' -1/2,-k_y -1/2}^{({\rm b})}$ are not equal to each other for circularly polarized radiation. Such a spin-dependent asymmetry of the optical transitions leads to a circular photocurrent.

The circular photocurrent driven by the indirect optical transitions can be estimated as 
\begin{equation}
\label{jdrude}
j_y^{\rm (CPGE)} \sim e \frac{d (v \tau_p)}{d \eps} (\eps_{F})  \left( w_{+1/2} - w_{-1/2} \right) I\:,
\end{equation}
where $w_s$ is the absorption width of the edge mode $s$. The difference of the absorption widths $w_{+1/2}$ and $w_{-1/2}$ is proportional to $P_{\rm circ}$ and can be presented in the form
\begin{equation}
w_{+1/2} - w_{-1/2} = w K' P_{\rm circ} o_z \:,
\end{equation}
where $w = w_{+1/2} + w_{-1/2}$ is the total absorption width of the edge channel
in this spectral range and $K'$ is a dimensionless coefficient.
This coefficient can be estimated as
$K' \sim (\hbar\omega/\delta_0) K$, where $K$ is defined by Eq.~\eqref{K}.

The absorption width for the indirect optical transitions can be estimated as 
\begin{equation}\label{w_Drude}
w \sim \frac{\A e^2}{c n_{\omega} \, \delta_0^2 \, \tau_p^{\rm bulk}} \:,
\end{equation}
where $\tau_p^{\rm bulk} \sim \A^2 \hbar/(\langle V_{E1}^2 \rangle \delta_0)$ is the momentum relaxation time of 2D carriers, $\langle V_{E1}^2 \rangle$ is the square of the matrix element of scattering averaged over impurities. 

Equations~\eqref{jdrude}-\eqref{w_Drude} provide the following estimation for the circular photocurrent caused by spin-conserving optical transitions in the edge channel
\begin{equation}\label{j_drude_spincons}
j_y^{\rm (CPGE)} \sim \frac{\A e^3 \hbar \omega }{c n_{\omega} \, \delta_0^3 \, \tau_p^{\rm bulk}} \frac{d (v \tau_p)}{d \eps} (\eps_{F})  I P_{\rm circ} o_z \:. 
\end{equation}
It shows that $j_y^{\rm (CPGE)}$ is about 10~pA for the radiation intensity $I = 1$~W/cm$^2$, 
the band gap $2|\delta_0| = 20$~meV, the photon energy $\hbar \omega = 1$~meV, the relaxation time $\tau_p^{\rm bulk} = 0.3$~ps (extracted from the mobility of bulk carriers~\cite{Dantscher2017}),  $d (v \tau_p) /d \eps \sim v_0 \tau_p/|\delta_0|$, and the relaxation time of edge carriers $\tau_p = 20$~ps.
 
(b) For indirect optical transitions between the spin branches, Fig.~\ref{fig:drude}b, the edge photocurrent can be presented 
in the form
\begin{equation}
\label{jdrude2}
j_y = \frac{2 e v_{0} \tau_p}{\hbar \omega} \left( w_{+1/2,-1/2} - w_{-1/2,+1/2} \right) I \:,
\end{equation}
where $w_{s,-s}$ are the effective absorption widths for the spin-flip transitions.  

The indirect transitions with spin flip can also occur via intermediate states in the conduction or valence  subbands of the QW. The difference of the absorption widths for this kind of processes can be estimated as
\begin{equation}\label{w_drude_spinflip}
w_{+1/2,-1/2} - w_{-1/2,+1/2} \sim \frac{\langle V_{E1,H1}^2 \rangle}{\langle V_{E1}^2 \rangle}  
\frac{(\hbar\omega)^2}{\delta_0^2} w \:,
\end{equation}
where $V_{E1,H1}$ and $V_{E1}$ are the matrix elements of inter-subband and intra-subband scattering, respectively, and $w$ is the absorption width given by Eq.~\eqref{w_Drude}. The photocurrent caused by spin-flip indirect optical transitions~\eqref{jdrude2} is order of magnitude smaller than the photocurrent caused by spin-conserving optical transitions~\eqref{j_drude_spincons}.

\subsection{Experimental results}

Edge photocurrents in 2D topological insulators were observed in Ref.~\cite{Dantscher2017}. 
The experiments were carried out on Hg$_{0.3}$Cd$_{0.7}$Te/HgTe/Hg$_{0.3}$Cd$_{0.7}$Te QW structures 
with the well width of 8 nm grown by molecular beam epitaxy on GaAs substrates. The photocurrent was excited 
by circularly polarized terahertz radiation with the photon energy smaller than the QW band gap. It was observed
that photocurrents on the opposite edges of the samples flowed consistently in the opposite directions.
The direction of the edge photocurrent was determined by the sign of photon helicity, i.e.,
the photocurrent was reversed by switching the radiation polarization from $\sigma^+$ to $\sigma^-$.
The samples were equipped with semitransparent top gates which enabled the tuning of the Fermi level.

It was observed that the edge photocurrent is efficiently generated when the Fermi level is in the band gap
and has a maximum at the Fermi level lying somewhat below the conduction-band bottom~\cite{Dantscher2017}.
Analysis of the dependence of the edge photocurrent on the gate voltage $V_g$ and the efficiency 
of different mechanisms of current generation revealed that
the photocurrent in this range of $V_g$ is dominated by photoionization of the edge states
 into the conduction band. The performed microscopic calculations supported this scenario.
 At negative $V_g$ where the Fermi level is shifted towards the valence band, the photocurrent
 changes its sign and exhibits another rise. This rise may be related to optical transitions from 
 the valence band to the edge states. The change of the photocurrent sign, however, is unclear
 and odds with the BHZ model which suggests the same current direction for the edge-to-conduction-band
 and valence-band-to-edge transitions~\cite{Dantscher2017,Magarill2016}. It may be related to the complex structure of the valence band in  8-nm-wide HgTe QWs not included in the BHZ model. Another explanation could be that the photocurrent
 in this range of $V_g$ is related to indirect optical transitions within the edge channel which are manifested
 due to strong energy dependence of the relaxation time or nonlinearity of the edge-state dispersion.
 
 The study of QW-based topological insulators with higher quality and larger band gap,
 which are already available~\cite{Leubner2016}, will enable the observation of edge photocurrents related 
 to different  types of optical transitions and provide valuable information about the 
 properties of helical states. 
 
 To summarize, we have discussed the high-frequency and optical properties of helical edge channels in two-dimensional
 topological insulators based on zinc-blende-type quantum wells. The interaction of electrons in the edge channel with a polarized electromagnetic wave is spin dependent and asymmetric in the momentum space, which results in edge photocurrents sensitive to the photon helicity. Depending on the radiation frequency and the position of the Fermi level in the structure, the
 edge photocurrents are contributed by different types of optical transitions. The most probable optical transitions involving edge states are direct transitions between the edge states and the bulk states in the quantum well. They are allowed already in centrosymmetric models on topological insulators and determine the edge photocurrent if the photon energy is large enough to  throw up edge electrons (or holes) to the bulk conduction-band (or valence-band) states. Optical transitions within the edge channel come into play for smaller photon energies. There can be direct transitions between the ``spin-up'' and ``spin-down'' edge states and indirect transitions assisted by electron scattering from static defects or phonons. The former occur in the electro-dipole approximation due to the lack of space inversion center in the quantum well while the latter require electron scattering. The relative
strength of the photocurrents contributed by these direct and indirect transitions depends on the quality and the band-structure parameters of the topological insulator.

The work was supported by the Russian Science Foundation (project 17-12-01265).


\begin{thebibliography}{000}


\bibitem{Hasan2010}
M. Z. Hasan and C. L. Kane,
Colloquium: Topological insulators,
Rev. Mod. Phys. \textbf{82}, 3045 (2010).

\bibitem{Zhang2011}
X.-L. Qi and S.-C. Zhang.
Topological insulators and superconductors,
Rev. Mod. Phys. \textbf{83}, 1057 (2011).

\bibitem{Moore2010} 
J. E. Moore,  
The birth of topological insulators, 
Nature \textbf{464}, 194 (2010).

\bibitem{Volkov1986} 
B. A. Volkov and O. A. Pankratov,
Inverted contact in semiconductors -- a new inhomogeneous structure with a 
two-dimensional gas of zero-mass electrons,
Sov. Phys. Uspekhi {\bf 29}, 579 (1986).

\bibitem{Volkov1985} 
B. A. Volkov and O. A. Pankratov, 
Two-dimensional massless electrons in an inverted contact,
JETP Lett. \textbf{42}, 178 (1985). 

\bibitem{Pankratov1987}
O. A. Pankratov, S. V. Pakhomov, and B. A. Volkov,
Supersymmetry in heterojunctions: Band-inverting contact on the basis 
of Pb$_{1-x}$Sn$_x$Te and Hg$_{1-x}$Cd$_x$Te,
Solid State Commun. {\bf 61}, 93 (1987).

\bibitem{Bernevig2006a}
B. A. Bernevig and S.-C. Zhang,
Quantum spin Hall effect,
Phys. Rev. Lett. {\bf 96}, 106802 (2006).

\bibitem{Kane2005a}
C. L. Kane and E. J. Mele,
$Z_2$ topological order and the quantum spin Hall effect,
Phys. Rev. Lett. {\bf 95}, 146802 (2005).

\bibitem{Fu2007a}
L. Fu, C. L. Kane, and E. J. Mele.  
Topological insulators in three dimensions, 
Phys. Rev. Lett. \textbf{98}, 106803 (2007).

\bibitem{Ando2013}
Y.  Ando,
Topological insulator materials, 
J. Phys. Soc. Japan \textbf{82}, 102001 (2013).


\bibitem{Zhang2009}
H. Zhang, C.-X. Liu, X.-L. Qi, X. Dai, Z. Fang, and S.-C. Zhang,
Topological insulators in Bi$_2$Se$_3$, Bi$_2$Te$_3$ and Sb$_2$Te$_3$ with a single Dirac cone on the surface, 
Nat. Phys. \textbf{5}, 438 (2009).

\bibitem{Xia2009}
Y. Xia, D. Qian, D. Hsieh, L. Wray, A. Pal, H. Lin, A. Bansil, D. Grauer, Y. S. Hor, R. J. Cava, and M. Z. Hasan,  
Observation of a large-gap topological-insulator class with a single Dirac cone on the surface, 
Nat. Phys. \textbf{5}, 398 (2009).

\bibitem{Fu2007b}
L. Fu and C. L. Kane,
Topological insulators with inversion symmetry,
Phys. Rev. B {\bf 76}, 045302 (2007).

\bibitem{Brune2011}
C. Br\"une, C. X. Liu, E. G. Novik, E. M. Hankiewicz, H. Buhmann, Y. L. Chen, X. L. Qi, Z. X. Shen, S.-C. Zhang, 
and L. W. Molenkamp,  
Quantum Hall effect from the topological surface states of strained bulk HgTe,
Phys. Rev. Lett. \textbf{106}, 126803 (2011).

\bibitem{Kozlov2014}
D. A. Kozlov, Z. D. Kvon, E. B. Olshanetsky, N. N. Mikhailov, S. A. Dvoretsky, and D. Weiss,  
Transport properties of a 3D topological insulator based on a strained high-mobility HgTe film, 
Phys. Rev. Lett. \textbf{112}, 196801 (2014).


\bibitem{Bernevig2006} 
B. A. Bernevig, T. L. Hughes, and S.-C. Zhang,
Quantum spin hall effect and topological phase transition in HgTe quantum wells,
Science \textbf{314}, 1757 (2006).

\bibitem{Konig2007}
M. K\"onig, S. Wiedmann, C. Br\"{u}ne, A. Roth, H. Buhmann, L. W. Molenkamp, X.-L. Qi, and S.-C. Zhang,  
Quantum spin Hall insulator state in HgTe quantum wells, 
Science \textbf{318}, 766 (2007).

\bibitem{Liu2008} 
C. Liu, T. L. Hughes, X.-L. Qi, K. Wang, and S.-C. Zhang,
Quantum spin Hall effect in inverted type-II semiconductors,
Phys. Rev. Lett. {\bf 100}, 236601 (2008).

\bibitem{Krishtopenko2018}
S. S. Krishtopenko, S. Ruffenach, F. Gonzalez-Posada, G. Boissier, M. Marcinkiewicz, M. A. Fadeev, A. M. Kadykov, 
V. V. Rumyantsev, S. V. Morozov, V. I. Gavrilenko, C. Consejo, W. Desrat, B. Jouault, W. Knap, E. Tournie, and F. Teppe,
Temperature-dependent terahertz spectroscopy of inverted-band three-layer InAs/GaSb/InAs quantum well,
Phys. Rev. B {\bf 97}, 245419 (2018).
 
\bibitem{Qian2014}
X. Qian, J. Liu, L. Fu, and J. Li, 
Quantum spin Hall effect in two-dimensional transition metal dichalcogenides, 
Science \textbf{20}, 1256815 (2014).

\bibitem{Fei2017}
Z. Fei, T. Palomaki, S. Wu, W. Zhao, X. Cai, B. Sun, P. Nguyen, J. Finney, X. Xu, and D. H. Cobden,  
Edge conduction in monolayer WTe2, 
Nat. Phys. \textbf{13}, 677 (2017).


\bibitem{Hsieh2009}
D. Hsieh, Y. Xia, D. Qian, L. Wray, J. H. Dil, F. Meier, J. Osterwalder, L. Patthey, J. G. Checkelsky, N. P. Ong, 
A. V. Fedorov, H. Lin, A. Bansil, D. Grauer, Y. S. Hor, R. J. Cava, and M. Z. Hasan,  
A tunable topological insulator in the spin helical Dirac transport regime, 
Nature \textbf{460}, 1101 (2009).

\bibitem{Checkelsky2009} 
J. G. Checkelsky, Y. S. Hor, M.-H. Liu, D.-X. Qu, R. J. Cava, and N. P. Ong,
Quantum interference in macroscopic crystals of nonmetallic Bi$_2$Se$_3$,
Phys. Rev. Lett. {\bf 103}, 246601 (2009).

\bibitem{Analytis2010}
J. G. Analytis, J.-H. Chu, Y. Chen. F. Corredor, R. D. McDonald, Z. X. Shen, and I. R. Fisher,
Bulk Fermi surface coexistence with Dirac surface state in Bi$_2$Se$_3$: A comparison of photoemission and Shubnikov-de-Haas measurements, 
Phys. Rev. B {\bf 81}, 205407 (2010).


\bibitem{Ren2011} 
Z. Ren, A. A. Taskin, S. Sasaki, K. Segawa, and Y. Ando,
Optimizing Bi$_{2-x}$Sb$_x$Te$_{3-y}$Se$y$ solid solutions to approach the intrinsic topological insulator regime,
Phys. Rev. B {\bf 84}, 165311 (2011).

\bibitem{Xia2013} 
B. Xia, P. Ren, A. Sulaev, P. Liu, S.-Q. Shen, and L. Wang,
Indications of surface-dominated transport in single crystalline nanoflake devices of topological insulator 
Bi$_{1.5}$Sb$_{0.5}$Te$_{1.8}$Se$_{1.2}$,
Phys. Rev. B {\bf 87}, 085442 (2013).

\bibitem{Pan2014} 
Y. Pan, D. Wu, J. R. Angevaare, H. Luigjes, E. Frantzeskakis, N. de Jong, E. van Heumen, T. V. Bay, 
B. Zwartsenberg, and Y. K. Huang,
Low carrier concentration crystals of the topological insulator Bi$_{2-x}$Sb$_x$Te$_{3-y}$Se$_y$: 
A magnetotransport study,
New J. Phys. {\bf 16}, 123035 (2014).

\bibitem{Shuvaev2013b} 
A. M. Shuvaev, G. V. Astakhov, G. Tkachov, C. Br\"{u}ne, H. Buhmann, L. W. Molenkamp, and A. Pimenov,
Terahertz quantum Hall effect of Dirac fermions in a topological insulator,
Phys. Rev. B {\bf 87}, 121104(R) (2013).

\bibitem{Dziom2017}
V. Dziom, A. Shuvaev, A. Pimenov, G. V. Astakhov, C. Ames, K. Bendias, J. B\"{o}ttcher, G. Tkachov, E. M. Hankiewicz, 
C. Br\"{u}ne, H. Buhmann, and L. W. Molenkamp,
Observation of the universal magnetoelectric effect in a 3D topological insulator,
Nat. Comm. {\bf 8}, 15197 (2017).


\bibitem{Roth2009}
A. Roth, C. Br\"{u}ne, H. Buhmann, L. W. Molenkamp, J. Maciejko, X.-L. Qi, and S.-C. Zhang, 
Nonlocal transport in the quantum spin Hall state, Science \textbf{325}, 294 (2009).

\bibitem{Gusev2011}
G. M. Gusev, Z. D. Kvon, O. A. Shegai, N. N. Mikhailov, S. A. Dvoretsky, and J. C. Portal,
Transport in disordered two-dimensional topological insulators,
Phys. Rev. B \textbf{84}, 121302(R) (2011).

\bibitem{Hart2014}
S. Hart, H. Ren, T. Wagner, P. Leubner, M. M\"uhlbauer, C. Br\"une, H. Buhmann, L. W. Molenkamp, and A. Yacoby,
Induced superconductivity in the quantum spin Hall edge, 
Nat. Phys. \textbf{10}, 638 (2014).

\bibitem{Ma2015} 
E. Y. Ma, M. R. Calvo, J. Wang, B. Lian, M. M\"{u}hlbauer, C. Br\"{u}ne, Y.-T. Cui, K. Lai, 
W. Kundhikanjana, Y. Yang, M. Baenninger, M. K\"{o}nig, C. Ames, H. Buhmann, P. Leubner, L. W. Molenkamp, S.-C. Zhang, 
D. Goldhaber-Gordon, M. A. Kelly, and Z.-X. Shen,
Unexpected edge conduction in mercury telluride quantum wells under broken time-reversal symmetry,
Nat. Comm. {\bf 6}, 7252 (2015).

\bibitem{Tikhonov2015} 
E. S. Tikhonov, D. V. Shovkun, V. S. Khrapai, Z. D. Kvon, N. N. Mikhailov, and
S. A. Dvoretsky, Shot noise of the edge transport in the inverted band HgTe quantum wells,
JETP Lett. \textbf{101}, 708 (2015).

\bibitem{Kononov2015} 
A. Kononov, S. V. Egorov, Z. D. Kvon, N. N. Mikhailov, S. A. Dvoretsky, and E. V. Deviatov,  
Evidence on the macroscopic length scale spin coherence for the edge currents in a narrow HgTe quantum well, 
JETP Lett. \textbf{101}, 814 (2015).

\bibitem{Kadykov2018} 
A. M. Kadykov, S. S. Krishtopenko, B. Jouault, W. Desrat, W. Knap, S. Ruffenach, C. Consejo, J. Torres, 
S. V. Morozov, N. N. Mikhailov, S. A. Dvoretskii, and F. Teppe,
Temperature-induced topological phase transition in HgTe quantum wells,
Phys. Rev. Lett. {\bf 120},  086401 (2018).


\bibitem{Magarill1979}
L. I. Maragill and M. V. Entin,
Photogalvanic effect in films,
Sov. Phys. Solid State {\bf 21}, 743 (1979).

\bibitem{Alperovich1982}
V. L. Alperovich, V. I. Belinicher, V. N. Novikov, and A. S. Terekhov,
Photogalvanic effects investigation in gallium arsenide, 
Ferroelectrics {\bf 45}, 1 (1982).

\bibitem{Schmidt2015}
C. B. Schmidt, S. Priyadarshi, S. A. Tarasenko, and M. Bieler, 
Ultrafast magneto-photocurrents in GaAs: Separation of surface and bulk contributions, 
Appl. Phys. Lett. {\bf 106}, 142108 (2015).

\bibitem{Karch2011}
J. Karch, C. Drexler, P. Olbrich, M. Fehrenbacger, M. Hirmer, M. M. Glazov, S. A. Tarasenko, E. L. Ivchenko, B. Birkner, J. Eroms, D. Weiss, R. Yakimova, S. Lara-Avila, S. Kubatkin, M. Ostler, T. Seyller, and S. D. Ganichev, 
Terahertz radiation driven chiral edge currents in graphene,
Phys. Rev. Lett. {\bf 107}, 276601 (2011).


\bibitem{Reimann2002}
P. Reimann, Brownian motors: noisy transport far from equilibrium,
Phys. Rep. {\bf 361}, 57 (2002).

\bibitem{Falko1989}
V. I. Fal'ko, 
Rectifying properties of 2D inversion layers in a parallel magnetic field, 
Sov. Phys. Solid State {\bf 31}, 561 (1989).

\bibitem{Tarasenko2008}
S. A. Tarasenko,
Direct current driven by ac electric field in quantum wells,
Phys. Rev. B {\bf  83}, 035313 (2011).

\bibitem{Drexler2013} C. Drexler, S. A. Tarasenko, P. Olbrich, J. Karch, M. Hirmer,
F. M\"{u}ller, M. Gmitra, J. Fabian, R. Yakimova, S. Lara-Avila,
S. Kubatkin, M. Wang, R. Vajtai, P. M. Ajayan, J. Kono, and
S. D. Ganichev, Magnetic quantum ratchet effect in graphene,
Nat. Nanotechnol. {\bf 8}, 104 (2013).


\bibitem{SturmanFridkin} B. I. Sturman and V. M. Fridkin, {\it The Photovoltaic and Photorefractive
Effects in Non-Centrosymmetric Materials} (Gordon and Breach, New York, 1992).

\bibitem{IvchenkoGanichev} E. L. Ivchenko and S. D. Ganichev, {\it Spin Photogalvanics in Spin
Physics in Semiconductors}, edited by M. I. Dyakonov (Springer, Berlin, 2008).

\bibitem{Glass1974} 
A.M. Glass, D. von der Linde, T.J. Negran, 
Appl. Phys. Lett.  (1974).

\bibitem{Asnin78} 
V. M. Asnin, A. A. Bakun,  A.M. Danishevskii, E.L. Ivchenko, G.E. Pikus, A.A. Rogachev,
Observation of a photo-emf that depends on the sign of the circular polarization of the light,
JETP Lett. {\bf 28}, 74 (1978).


\bibitem{Ganichev2003} S. D. Ganichev, V. V. Bel'kov, P. Schneider, E. L. Ivchenko, S. A. Tarasenko, 
W. Wegscheider, D. Weiss, D. Schuh, E. V. Beregulin, and W. Prettl, 
Resonant inversion of the circular photogalvanic effect in $n$-doped quantum wells, 
Phys. Rev. B {\bf 68}, 035319 (2003).

\bibitem{Bieler2005}
M. Bieler, N. Laman, H. M. van Driel, and A. L. Smirl,
Ultrafast spin-polarized electrical currents injected in a strained zinc blende semiconductor by single color pulses,
Appl. Phys. Lett. {\bf 86}, 061102 (2005).

\bibitem{Yang2006}
C. L. Yang, H. T. He, Lu Ding, L. J. Cui, Y. P. Zeng, J. N. Wang, and W. K. Ge,
Spectral dependence of spin photocurrent and current-induced spin polarization 
in an InGaAs/InAlAs two-dimensional electron gas,
Phys. Rev. Lett. {\bf 96}, 186605 (2006).

\bibitem{Olbrich2009}
P. Olbrich, J. Allerdings, V. V. Bel'kov, S. A. Tarasenko, D. Schuh, W. Wegscheider, T. Korn, C. Sch\"{u}ller,
D. Weiss, and S. D. Ganichev, 
Magnetogyrotropic photogalvanic effect and spin dephasing in (110)-grown GaAs/AlGaAs quantum well structures, 
Phys. Rev. B {\bf 79}, 245329 (2009).

\bibitem{Priyadarshi2012}
S. Priyadarshi, K. Pierz, and M. Bieler, 
All-optically induced ultrafast photocurrents: Beyond the instantaneous coherent response,
Phys. Rev. Lett. {\bf 109}, 216601 (2012).

\bibitem{Olbrich2013}
P. Olbrich, C. Zoth, P. Vierling, K.-M. Dantscher, G. V. Budkin, S. A. Tarasenko, V. V. Bel'kov, D. A. Kozlov, Z. D. Kvon, 
N. N. Mikhailov, S. A. Dvoretsky, and S. D. Ganichev, 
Giant photocurrents in a Dirac fermion system at cyclotron resonance, 
Phys. Rev. B {\bf 87}, 235439 (2013).

\bibitem{Ganichev2014}
S. D. Ganichev and L. E. Golub,
Interplay of Rashba/Dresselhaus spin splittings probed by photogalvanic spectroscopy -- A review,
Phys. Status Solidi B {\bf 251}, 1801 (2014).

\bibitem{Li2017} 
J. Li, W. Yang, J.-T. Liu, W. Huang, C. Li, and S.-Y. Chen,
Enhanced circular photogalvanic effect in HgTe quantum wells in the heavily inverted regime,
Phys. Rev. B {\bf 95}, 035308 (2017).

\bibitem{Mikheev2018}
G. M. Mikheev, A. S. Saushin, V. M. Styapshin, and Yu. P. Svirko,
Interplay of the photon drag and the surface photogalvanic effects in the metal-semiconductor nanocomposite,
Sci. Rep. {\bf 8}, 8644 (2018).
 

\bibitem{Bhat2005} 
R. D. R. Bhat, F. Nastos, N. Ali, and J. E. Sipe,
Pure spin current from one-photon absorption of linearly polarized light in noncentrosymmetric semiconductors,
Phys. Rev. Lett. {\bf 94}, 96603 (2005).

\bibitem{Tarasenko2005}
S.A. Tarasenko and E.L. Ivchenko, 
Pure spin photocurrents in low-dimensional structures, 
JETP Lett. {\bf 81}, 231 (2005).

\bibitem{Zhao2005} 
H. Zhao, X. Pan, A. L. Smirl, R. D. R. Bhat, A. Najmaie, J. E. Sipe, and H. M. van Driel, 
Injection of ballistic pure spin currents in semiconductors by a single-color linearly polarized beam,
Phys. Rev. B {\bf 72}, 201302 (2005).

\bibitem{Ganichev2006} S. D. Ganichev, V. V. Bel'kov, S. A. Tarasenko, S. N. Danilov, S. Giglberger, Ch. Hoffmann, 
E. L. Ivchenko, D. Weiss, W. Wegscheider, C. Gerl, D. Schuh, J. Stahl, J. De Boeck, G. Borghs, and W. Prettl, 
Zero-bias spin separation, 
Nat. Phys. {\bf 2}, 609 (2006).  

\bibitem{Ganichev2009} S. D. Ganichev, S. A. Tarasenko, V. V. Bel'kov, P. Olbrich, W. Eder, D.R. Yakovlev, V. Kolkovsky, 
W. Zaleszczyk, G. Karczewski, T. Wojtowicz, and D. Weiss, 
Spin currents in diluted magnetic semiconductors, 
Phys. Rev. Lett. {\bf 102}, 156602 (2009). 


\bibitem{Karch2011b}
J. Karch, S. A. Tarasenko, E. L. Ivchenko, J. Kamann, P. Olbrich, M. Utz, Z. D. Kvon, S. D. Ganichev, 
Photoexcitation of valley-orbit currents in (111)-oriented silicon metal-oxide-semiconductor field-effect transistors, 
Phys. Rev. B {\bf 83}, 121312(R) (2011).

\bibitem{Golub2011} 
L. E. Golub, S. A.  Tarasenko, M. V.  Entin, and L. I.  Magarill, 
Valley separation in graphene by polarized light, 
Phys. Rev. B {\bf 84}, 195408 (2011).

\bibitem{Hartmann2011} 
R. R. Hartmann and M. E. Portnoi, 
{\it Optoelectronic Properties of Carbon-based Nanostructures: Steering electrons in graphene
by electromagnetic fields} (LAP LAMBERT Academic Publishing,
Saarbrucken, 2011).

\bibitem{Yuan2014}
H. Yuan, X. Wang, B. Lian, H. Zhang, X. Fang, B. Shen, G. Xu, Y. Xu, Sh.-Ch. Zhang, H. Y. Hwang,
and Y. Cui, Generation and electric control of spin-valley-coupled circular photogalvanic current in WSe$_2$,
Nat. Nanotechnol. {\bf 9}, 851 (2014).

\bibitem{Linnik2014}
T. L. Linnik,
Photoinduced valley currents in strained graphene,
Phys. Rev. B {\bf 90}, 075406 (2014).

\bibitem{LyandaGeller2015}
Y. B. Lyanda-Geller, S. Li, and A. V. Andreev,
Polarization-dependent photocurrents in polar stacks of van der Waals solids,
Phys. Rev. B {\bf 92}, 241406(R) (2015).


\bibitem{Hsieh2011} 
D. Hsieh, J. W. McIver, D. H. Torchinsky, D. R. Gardner, Y. S. Lee, and N. Gedik,
Nonlinear optical probe of tunable surface electrons on a topological insulator,
Phys. Rev. Lett. {\bf 106}, 057401 (2011).

\bibitem{McIver2012}
J. W. McIver, D. Hsieh, H. Steinberg, P. Jarillo-Herrero, and N. Gedik,
Control over topological insulator photocurrents with light polarization,
Nat. Nanotech. {\bf 7}, 96 (2012).

\bibitem{Olbrich2014} 
P. Olbrich, L. E. Golub, T. Herrmann, S. N. Danilov, H. Plank, V. V. Bel'kov, G. Mussler, Ch. Weyrich, 
C. M. Schneider, J. Kampmeier, D. Gr\"{u}tzmacher, L. Plucinski, M. Eschbach, and S. D. Ganichev,
Room-temperature high-frequency transport of Dirac fermions in epitaxially grown 
Sb$_2$Te$_3$- and Bi$_2$Te$_3$-based topological insulators,
Phys. Rev. Lett. {\bf 113}, 096601 (2014).

\bibitem{Dantscher2015} 
K.-M. Dantscher, D. A. Kozlov, P. Olbrich, C. Zoth, P. Faltermeier, M. Lindner, G. V. Budkin, S. A. Tarasenko, 
V. V. Bel'kov, Z. D. Kvon, N. N. Mikhailov, S. A. Dvoretsky, D. Weiss, B. Jenichen, and S. D. Ganichev,
Cyclotron-resonance-assisted photocurrents in surface states of a three-dimensional topological insulator 
based on a strained high-mobility HgTe film,
Phys. Rev. B {\bf 92}, 165314 (2015).

\bibitem{Shikin2016} 
A. M. Shikin, I. Klimovskikh, M. V. Filyanina, A. A. Rybkina, D. A. Pudikov, K. A. Kokh, and O. E. Tereshchenko,
Surface spin-polarized currents generated in topological insulators by circularly polarized synchrotron radiation and their 
photoelectron spectroscopy indication,
Phys. Solid State {\bf 58}, 1675 (2016).

\bibitem{Hamh2016} 
S. Y. Hamh, S.-H. Park, S.-K. Jerng, J. H. Jeon, S.-H. Chun, and J. S. Lee,
Helicity-dependent photocurrent in a Bi$_2$Se$_3$ thin film probed by terahertz emission spectroscopy,
Phys. Rev. B {\bf 94}, 161405(R) (2016). 

\bibitem{Galeeva2016} 
A. V. Galeeva, S. G. Egorova, V. I. Chernichkin, M. E. Tamm, L. V. Yashina, V. V. Rumyantsev, S. V. Morozov, 
H. Plank, S. N. Danilov, L. I. Ryabova, and D. R. Khokhlov,
Manifestation of topological surface electron states in the photoelectromagnetic effect induced by terahertz laser radiation,
Semicond. Sci. Technol. {\bf 31}, 095010 (2016).

\bibitem{Pan2017} 
Y. Pan, Q.-Z. Wang, A. L. Yeats, T. Pillsbury, T. C. Flanagan, A. Richardella, H. Zhang, D. D. Awschalom, C.-X. Liu, and
N. Samarth,
Helicity dependent photocurrent in electrically gated (Bi$_{1-x}$Sb$_x$)$_2$Te$_3$ thin films,
Nat. Comm. {\bf 8}, 1037 (2017).

\bibitem{Huang2017} 
Y. Q. Huang, Y. X. Song, S. M. Wang, I. A. Buyanova, and W. M. Chen,
Spin injection and helicity control of surface spin photocurrent in a three dimensional topological insulator,
Nat. Comm. {\bf 8}, 15401 (2017).

\bibitem{Kuroda2017} 
K. Kuroda, J. Reimann, K. A. Kokh, O. E. Tereshchenko, A. Kimura, J. G\"{u}dde, and U. H\"{o}fer,
Ultrafast energy- and momentum-resolved surface Dirac photocurrents in the topological insulator Sb$_2$Te$_3$,
Phys. Rev. B {\bf 95}, 081103(R) (2017).

\bibitem{Dantscher2017} 
K.-M. Dantscher, D. A. Kozlov, M. T. Scherr, S. Gebert, J. B\"{a}renf\"{a}nger, M. V. Durnev, 
S. A. Tarasenko, V. V. Bel'kov, N. N. Mikhailov, S. A. Dvoretsky, Z. D. Kvon, J. Ziegler, D. Weiss, and S. D. Ganichev, 
Photogalvanic probing of helical edge channels in two-dimensional HgTe topological insulators,	     
Phys. Rev. B {\bf 95}, 201103(R) (2017).

\bibitem{Plank2018}
H. Plank and S.D. Ganichev,
A review on terahertz photogalvanic spectroscopy of Bi$_2$Te$_3$- and Sb$_2$Te$_3$-based three dimensional 
topological insulators,
Solid-State Electronics {\bf 147}, 44 (2018). 


\bibitem{Barlow1954} 
H. M. Barlow,
Application of the Hall effect in a semiconductor to the measurement of power in an electromagnetic field,
Nature (London) {\bf 173}, 41 (1954).

\bibitem{Perel1973}
 V. I. Perel' and Ya. M. Pinskii, 
Constant current in conducting media due to a high-frequency electromagnetic field,
Sov. Phys. Solid State {\bf 15}, 688 (1973).

\bibitem{Shalygin2006}
V. A. Shalygin, H. Diehl, Ch. Hoffmann, S. N. Danilov, T. Herrle, S. A. Tarasenko, D. Schuh, Ch. Gerl, W. Wegscheider, 
W. Prettl, and S. D. Ganichev, 
Spin photocurrents and circular photon drag effect in (110)-grown quantum well structures, 
JETP Lett. {\bf 84}, 570 (2006).

\bibitem{Karch2010} 
J. Karch, J. Karch, P. Olbrich, M. Schmalzbauer, C. Zoth, C. Brinsteiner, M. Fehrenbacher, U. Wurstbauer, M. M. Glazov, 
S. A. Tarasenko, E. L. Ivchenko, D. Weiss, J. Eroms, R. Yakimova, S. Lara-Avila, S. Kubatkin, and S. D. Ganichev, 
Dynamic Hall effect driven by circularly polarized light in a graphene layer, 
Phys. Rev. Lett. {\bf 105}, 227402 (2010).

\bibitem{Obraztsov2014}
P. A. Obraztsov, N. Kanda, K. Konishi, M. Kuwata-Gonokami, S. V. Garnov, A. N. Obraztsov, and Yu. P. Svirko, 
Photon-drag-induced terahertz emission from graphene,
Phys. Rev. B {\bf 90}, 241416 (2014).

\bibitem{Shalygin2016}
V. A. Shalygin, M. D. Moldavskaya, S. N. Danilov, I. I. Farbshtein, L. E. Golub,
Circular photon drag effect in bulk tellurium,
Phys. Rev. B {\bf 93}, 045207 (2016).


\bibitem{Durnev2016}
M. V. Durnev and S. A. Tarasenko,
Magnetic field effects on edge and bulk states in topological insulators based on HgTe/CdHgTe quantum wells with strong
natural interface inversion asymmetry, 
Phys. Rev. B \textbf{93}, 075434 (2016). 

\bibitem{Tarasenko2015}
S. A. Tarasenko, M. V. Durnev, M. O. Nestoklon, E. L. Ivchenko, J.-W. Luo, and A. Zunger,  
Split Dirac cones in HgTe/CdTe quantum wells due to symmetry-enforced level anticrossing at interfaces, 
Phys. Rev. B \textbf{91}, 081302 (2015).

\bibitem{Dai2008}
X. Dai, T. L. Hughes, X.-L. Qi, Z. Fang, and S.-C. Zhang,
Helical edge and surface states in HgTe quantum wells and bulk insulators, 
Phys. Rev. B {\bf 77}, 125319 (2008).

\bibitem{Konig2008}
M. K\"onig, H. Buhmann, L. W. Molenkamp, T. Hughes, C.-X. Liu, X.-L. Qi, and S.-C. Zhang, 
The quantum spin Hall effect: Theory and experiment, J. Phys. Soc. Jpn. \textbf{77}, 031007 (2008).

\bibitem{Winkler2012} R. Winkler, L. Y. Wang, Y. H. Lin, and C. S. Chu,
Robust level coincidences in the subband structure of quasi-2D systems,
Solid State Commun. {\bf 152}, 2096 (2012).

\bibitem{Sonin2010}
E. B. Sonin, 
Edge accumulation and currents of moment in two-dimensional topological insulators, 
Phys. Rev. B {\bf 82}, 113307 (2010).

\bibitem{Scharf2012}
B. Scharf, A. Matos-Abiague, and J. Fabian, 
Magnetic properties of HgTe quantum wells, 
Phys. Rev. B {\bf 86}, 075418 (2012).

\bibitem{Klipstein2015} 
P. C. Klipstein, 
Structure of the quantum spin Hall states in HgTe/CdTe and InAs/GaSb/AlSb quantum wells, 
Phys. Rev. B {\bf 91}, 035310 (2015).

\bibitem{Enaldiev2015} 
V. V. Enaldiev, I. V. Zagorodnev, and V. A. Volkov, 
Boundary conditions and surface state spectra in topological insulators,
JETP Lett. {\bf 101}, 89 (2015).

\bibitem{Raichev2012} 
O. E. Raichev, 
Effective Hamiltonian, energy spectrum, and phase transition induced by in-plane magnetic field in symmetric
HgTe quantum wells, 
Phys. Rev. B {\bf 85}, 045310 (2012).

\bibitem{Krishtopenko2018}
S. S. Krishtopenko and F. Teppe,
Realistic picture of helical edge states in HgTe quantum wells, 
Phys. Rev. B {\bf 97}, 165408 (2018).

\bibitem{Minkov2016}
G. M. Minkov, A.V. Germanenko, O. E. Rut, A. A. Sherstobitov, M. O. Nestoklon, S. A. Dvoretski, and N. N. Mikhailov, 
Spin-orbit splitting of valence and conduction bands in HgTe quantum wells near the Dirac point,
Phys. Rev. B {\bf 93}, 155304 (2016).


\bibitem{Dora2012}
B. D\'ora, J. Cayssol, F. Simon, and R. Moessner, 
Optically engineering the topological properties of a spin Hall insulator, 
Phys. Rev. Lett. \textbf{108}, 056602 (2012).

\bibitem{Artemenko2013}
S. N. Artemenko and V. O. Kaladzhyan,
Photogalvanic effects in topological insulators, 
JETP Letters \textbf{97}, 82 (2013).

\bibitem{Durnev2018}
M. V. Durnev and S. A. Tarasenko,
Optical properties of helical edge channels in zinc-blende-type topological insulators: 
Selection rules, circular and linear dichroism, circular and linear photocurrents,
Journ. of Phys.: Cond. Matt. {\bf 31}, 035301 (2019).

\bibitem{Kaladzhyan2015}
V. Kaladzhyan, P. P. Aseev, and S. N. Artemenko, 
Photogalvanic effect in the HgTe/CdTe topological insulator due to edge-bulk optical transitions,
Phys. Rev. B {\bf 92}, 155424 (2015).

\bibitem{Magarill2016} 
L. I. Magarill and M. V. Entin, 
Circular photogalvanic effect caused by the transitions between edge and 2D states in a 2D topological insulator,
JETP Lett. {\bf 104}, 771 (2016).

\bibitem{Entin2016}
M. V. Entin and L. I. Magarill,
Edge absorption and circular photogalvanic effect in 2D topological insulator edges,
JETP Lett. {\bf 103}, 711 (2016).

\bibitem{Tarasenko2007} 
S. A. Tarasenko,
Orbital mechanism of the circular photogalvanic effect in quantum wells,
JETP Lett. \textbf{85}, 182 (2007).


\bibitem{Leubner2016} 
P. Leubner, L. Lunczer, C. Br\"une, H. Buhmann, and L. W. Molenkamp,  
Strain engineering of the band gap of HgTe quantum wells using superlattice virtual substrates, 
Phys. Rev. Lett. {\bf 117}, 086403 (2016).
    
\end{thebibliography}

\end{document}